\documentclass[twocolumn,showpacs,preprintnumbers,amsmath,amssymb]{revtex4}
\usepackage{graphicx}
\usepackage{dcolumn}
\usepackage{bm}
\usepackage{amssymb}
\usepackage{amsmath}
\begin{document}
\title{Electric--field effect on electron-doped infinite-layer Sr$_{0.88}$La$_{0.12}$CuO$_{2+x}$ thin films}

\author{L. Fruchter, F. Bouquet and Z.Z. Li}%
\affiliation{Laboratoire de Physique des Solides, Univ. Paris-Sud, CNRS, UMR 8502, F-91405 Orsay Cedex, France}
\date{Received: date / Revised version: date}

\begin{abstract}{We have used the electric--field effect to modulate the resistivity of the surface of underdoped Sr$_{0.88}$La$_{0.12}$CuO$_{2+x}$ thin films, allowing opposite modifications of the electron and hole density in the CuO$_2$ planes, an original situation with respect to conventional chemical doping in electron-doped materials. When the Hall effect indicates a large contribution of a hole band, the electric--field effect on the normal state resistivity is however dominated by the electrons, and the superconducting transition temperature increases when carriers are transfered from holes to electrons.}
\end{abstract}


\maketitle

It has long been known that the so--called electron-doped cuprates actually involve both electron and hole bands. This was first evidenced from Hall measurements, which generally exhibit both positive and negative contributions as a function of temperature\cite{Jiang1994,Budhani2002,Dagan2004,Jovanovic2009}. ARPES measurements on Nd$_{2-x}$Ce$_x$CuO$_4$ later allowed to identify the origin of these two contributions\cite{Armitage2001,Armitage2002}: doping the half--filled Mott insulator first introduces a (0,$\pi$) electron pocket. Further doping introduces a ($\pi/2$,$\pi/2$) hole pocket that coexists with the electron one, until they coalesce into a large ($\pi$,$\pi$) centered pocket. The resolution of a model t-t'-t"-U Hubbard Hamiltonian with appropriate choice of the band parameter allowed to reproduce and interpret the ARPES results: the electron and hole pockets are found to be respectively the upper and lower Hubbard bands of this model and, upon doping, the Fermi energy first moves into the upper band and then reaches the bottom of the hole band, as the Mott pseudogap shrinks\cite{Kusko2002}. The introduction of phenomenological d--wave superconductivity in this model allowed to account for the evolution from nodeless to d--wave like superfluid density as doping increases\cite{Das2008}, as observed in Ref.~\onlinecite{Fruchter2010}, as well as for the non--monotonic angular variation of the superconducting gap amplitude\cite{Matsui2005,Das2006}. Within this description, electron-doped cuprates are two--band superconductors (electron and hole) and it has been proposed that superconductivity in these bands is intrinsically coupled as a result of the antiferromagnetic fluctuations\cite{Das2006}.

The contribution of the electron band to superconductivity may be, however, questioned in several ways. First, in Nd$_{2-x}$Ce$_x$CuO$_4$, the optimum superconducting temperature is not reached until the hole pocket is well formed. Then, as noticed by Dagan et al\cite{Dagan2007}, in--plane resistivity for Pr$_{2-x}$Ce$_x$CuO$_4$ appears to be essentially determined by the electrons in a large doping range, including insulating states, and to be independent of the occurrence of superconductivity, both behaviors in sharp contrast with the Hall resistivity. This suggests that the hole pocket actually control the occurrence of superconductivity. Finally, it was shown, in the case of hole-doped cuprate, that quasi-particles resulting from the breaking of the Cooper pairs manifest themselves primarily in the vicinity of the nodal ($\pi/2$,$\pi/2$) direction of the d--wave superconducting gap\cite{Kohsaka2008}. This is also the direction along which the Fermi surface first appears in the pseudogap regime of underdoped hole cuprates\cite{Norman1998}. A similar behavior in electron-doped materials would imply that a significant Fermi surface also develops along the nodal direction, in order to develop the d-wave superconducting gap: this would be provided by the formation of the hole pocket.

As may be seen from the results of the Hubbard model, chemical doping simultaneously shifts the upper (electron) band and the lower (hole) band to the Fermi level so that, when both bands cross this level, there is an increase of both the electron and the hole densities in the CuO$_2$ planes with doping. As a result, it difficult to separate the contribution of these two bands to superconductivity. On the other hand, the electric--field effect allows the tuning of electron and hole densities in opposite ways, simply because of the opposite charges (see e.g. Ref. ~\onlinecite{Novoselov2004}). We have used this effect to tune the carrier density at the surface of electron-doped infinite-layer Sr$_{0.88}$La$_{0.12}$CuO$_{2+x}$ thin films. There is much less information on this compound than for the popular Nd$_{2-x}$Ce$_x$CuO$_4$ and Pr$_{2-x}$Ce$_x$CuO$_4$ materials (for a review, see Ref.~\onlinecite{Armitage2010}); however, similarities in the transport properties suggest that the evolution of the Fermi surface might be similar to what is described above, with a possible shift in the doping range\cite{Fruchter2010}.

The Thomas-Fermi length for the electrostatic screening of the gate potential is of the order of k$_F^{-1}$, and such a short length usually requires the use of ultra-thin samples, in order to achieve a uniform and sizeable electric--field effect (see e.g. Ref.~\onlinecite{Ahn2006}). Growth of ultrathin films invariably yields a degradation of the materials properties, which is not fully counterbalanced by the use of buffers between substrate and film. Our samples are also strongly constrained by the substrate\cite{Fruchter2010} and ultrathin samples cannot be grown from the pair substrate/material that we use to grow thick films. However, the electric--field effect on cuprates was originally investigated by doping the very first unit cells of thick films\cite{Fiory1990}. The simplicity of such a technique has several disadvantages. First, the transport properties cannot be measured independently of the underlying bulk material. The contribution of the investigated surface is very small, or may even be completely obscured when the bulk is superconducting. Moreover, the properties of the surface may differ strongly from that of the bulk, either due to the growing process or to chemical degradation. As we shall see, the latter inconvenience is limited in our case, as far as one can tell from nearly identical superconducting transition temperatures for the bulk film and the investigated surface.

The first inconvenience requires careful measurements, in order to be able to measure resistance variations that are typically 10$^{-5}$ of the total signal. We have used two 500 \AA ~thick films grown on a (100) KTaO$_3$ substrate, which have been patterned in the standard 4-wires resistivity measurement configuration. A 20 $\mu$m thick polymer film was laid on the resistivity bar (1000 x 350 $\mu$m$^2$), with a metallic electrode on the top of it. We have checked that the relative permittivity of the polymer film was close to unity in the whole temperature range, yielding a capacity $C = 2.3\, 10^{-13}$ Fd. In the vicinity of the superconducting transition, unavoidable temperature fluctuations (of the order of 10$^{-4}$) have a large effect on the measured resistance and standard DC measurements of the electric--field effect are not possible. We used the AC lock-in technique, modulating the gate voltage. Neither can large frequencies be used, as the current through the capacitor is proportional to this frequency and induces an out-of-phase signal (due to the asymmetry of the setup) which quickly overcomes the in-phase signal. A 1 Hz gate voltage with 100 V amplitude was found to be a good compromise. Possible electrostatic pressure effects on the film were intrinsically rejected from the measurements, such effects being an even function of the gate voltage, while the electric--field effect is an odd one. In order to eliminate the possibility of any experimental artifacts for these delicate measurements, we checked that the AC measurements results were identical to the conventional DC ones, in the normal state for selected temperatures.

For a simple modelling, we assume two bands of ideal free electron and hole gas. The electrostatic potential is screened at the scale of the Thomas-Fermi length given, in the case of weakly coupled conducting planes, by: $\lambda_{TF} = (\epsilon b / e^2 \partial n/\partial \mu)^{1/2}$, where $b$ is the CuO$_2$ plane separation, $\epsilon$ is the dielectric constant of the material, and $n$ is the surface density of carriers\cite{Ahn2006}. In the case of a 2D ideal free carrier gas, one has for the chemical potential $\mu \equiv \epsilon_F$ and $\partial n / \partial \mu = m/\pi \bar{h}^2$. This may be easily generalized to the case of two bands, using the replacement $\partial n/\partial \mu \rightarrow \partial n_e/\partial \mu - \partial n_h/\partial \mu$, where we adopt, by convention, $\partial n_e/\partial \mu > 0$ and $\partial n_h/\partial \mu < 0$. Assuming uniform electron and hole densities over this modified Thomas-Fermi length, one may straightforwardly generalize the results of Ref.~\onlinecite{Fiory1990} for the change in the sample sheet resistance, $\delta R$, using the parallel-resistance model:

\begin{equation}
\delta R/\delta Q = -C^{-1} \delta R/\delta V_G = - R^2 \, \frac{\delta Q_e \, \mu_e + \delta Q_h \, \mu_h}{\delta Q_h - \delta Q_e}
\label{eq1}
\end{equation}

where $\delta_Q$ is the total surfacic charge density induced at the sample surface, $C$ is the surfacic capacitance determined by the dielectric spacer, $\delta Q_e$ and $\delta Q_h$ are the electron and hole charge density variation, and $\mu_e$, $\mu_h$ the electron and hole mobilities. Making $\delta Q_e = 0$ in Eq.~\ref{eq1} shows that the resistance change for a single hole (electron) band material is a direct measure of the carrier mobility\cite{Fiory1990}: $\mu_{h,e} = \mp R^{-2}\delta R/\delta Q$. In the general case, it is shown easily that the change of hole (electron) charge density in the screening layer is:

\begin{equation}
\delta Q_{h,e} = \frac{\partial n_{e,h}/\partial \mu}{\partial n_{e}/\partial \mu-\partial n_{h}/\partial \mu} \delta Q
\label{eq2}
\end{equation}

where it is seen that a positive gate voltage induces an increase in the electron density and a depletion of the hole one. Combined with Eq.\ref{eq1}, this yields:

\begin{equation}
\delta R/\delta Q = R^2 \frac{\partial n_e/\partial \mu \,\mu_e +\partial n_h/\partial \mu  \,\mu_h}{\partial n_e/\partial \mu-\partial n_h/\partial \mu}
\label{eq3}
\end{equation}

which reduces in the case of ideal free carrier gas to:

\begin{equation}
\delta R/\delta Q = R^2 \frac{m_e \mu_e - m_h \mu_h}{m_e+m_h}= e\,R^2 \frac{\tau_e-\tau_h}{m_e+m_h}
\label{eq4}
\end{equation}

Thus, within the simple two--bands model, the electric--field effect provides an information complementary to the Hall coefficient:

\begin{equation}
R_H = \frac{n_h \mu_h^2-n_e \mu_e^2}{e (n_e\mu_e+n_h\mu_h)^2}
\label{eq5}
\end{equation}

and to the conductivity:

\begin{equation}
\sigma = e (n_e \mu_e + n_h \mu_h)
\label{eq6}
\end{equation}

\begin{figure}
\includegraphics[width= \columnwidth]{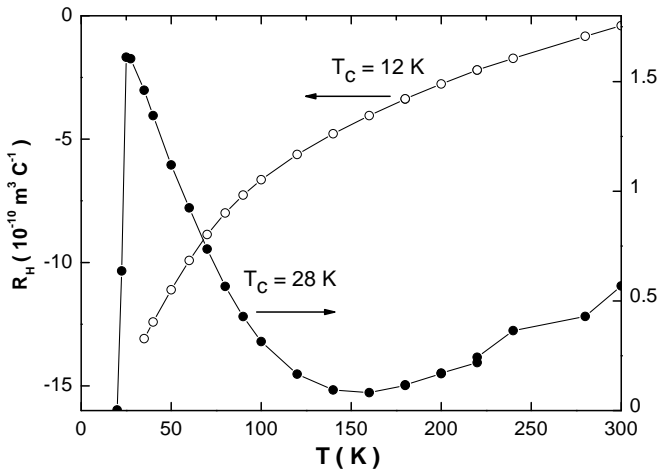}
\caption{Hall coefficient for the two Sr$_{0.88}$La$_{0.12}$CuO$_{2+x}$ thin films.}\label{champ0}
\end{figure}

\begin{figure}
\includegraphics[width= \columnwidth]{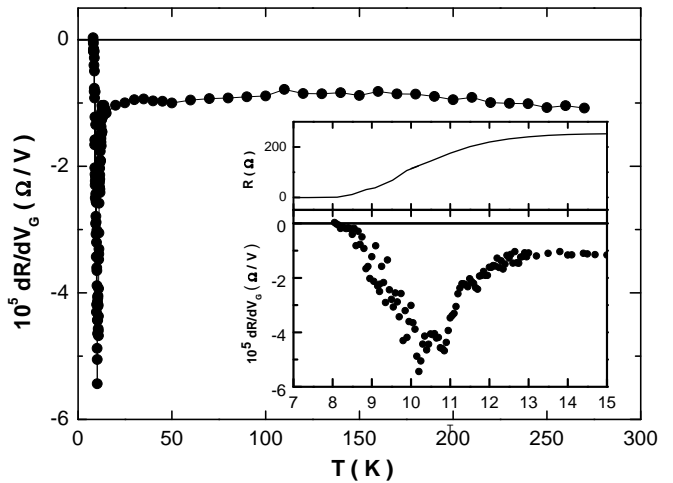}
\caption{Electric--field effect for the film with $T_c \approx$ 12 K. The inset shows the bulk (top) and surface (bottom) superconducting transition in detail.}\label{champ3}
\end{figure}

\begin{figure}
\includegraphics[width= \columnwidth]{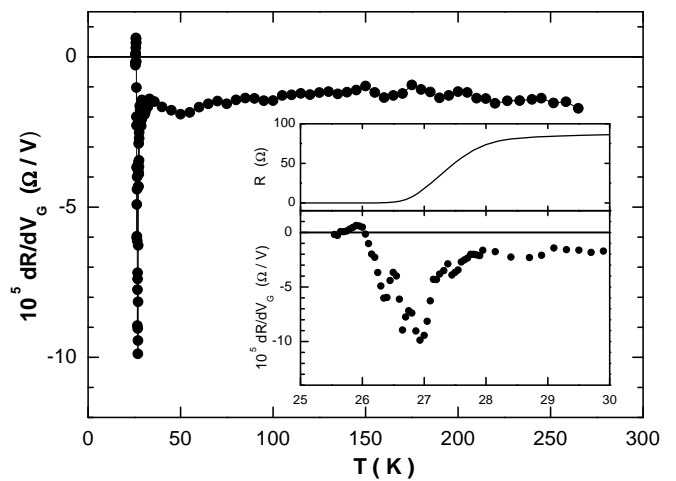}
\caption{Same as in Fig.~\ref{champ3} for the film with $T_c \approx$ 28 K}\label{champ1}
\end{figure}

Figs.~\ref{champ0}-\ref{champ1} display the Hall coefficient (independent of the magnetic field, in the range explored) and the electric--field--induced resistance change, measured for two Sr$_{0.88}$La$_{0.12}$CuO$_{2+x}$ thin films.  Doping in these samples was achieved both by Sr/La substitution and oxygen reduction, which is obtained by removing the apical oxygen after deposition\cite{Shin2001,Zli2009,Jovanovic2009}. The Hall angle for the less doped sample ($T_c \approx$ 12 K) is negative in the whole temperature range, whereas it is positive for the higher doping ($T_c \approx$ 28 K), a clear sign of a dominant hole contribution in this case. For both samples, the electric--field effect in the normal state is negative, thus revealing a dominant $\partial n_e/\partial \mu \,\mu_e$ term in Eq.~\ref{eq3}. We cannot, however, further extract carrier mobilities from Eqs.~\ref{eq3},\ref{eq5},\ref{eq6}, without further simplifying assumptions on their electron and hole terms.
 
We now focus on the results in a temperature range close to $T_c$. First, comparable temperatures for the onset of the superconducting transition in the bulk of the films and at the surface provide evidence that the electronic properties of the latter (at least, the one at the origin of the electric--field effect) are not strongly modified with respect to that of the former. Then, the large negative peak in $\delta R/\delta V_G$ is a clear indication that a simultaneous \textit{increase} of the electron density and \textit{decrease} of the hole one yield an \textit{increase} in $T_c$. As a consequence, electrons turn out to be essential to the occurrence of superconductivity for both dopings, in contradiction with previous arguments. There are several mechanisms which possibly account for our results.

First, it has been argued that a quantum critical point (QCP) is present near optimal doping in electron--doped materials, and that the quantum phase transition at this point is a magnetic one. Indeed, in the case of Pr$_{2-x}$Ce$_x$CuO$_4$, the low temperature resistivity and Hall coefficient under strong magnetic field ($T <$ 20 K) bear the signature of a QCP for a critical doping close to the optimal one \cite{Dagan2004}. It was proposed that this QCP corresponds to the merging of small electron and hole pockets into a large one, and to the concomitant vanishing of the Neel state at $T$ = 0 (see Ref.~\onlinecite{Kusko2002}, and Ref.~\onlinecite{Das2008} for the phenomenological introduction of d-wave superconductivity). In Ref.~\onlinecite{Moon2009} argues that this critical point is actually shifted to lower doping, in the presence of superconductivity (that is, in absence of a magnetic field). Within this scenario, the ground state at low doping is, in zero magnetic field, a superconductor coexisting with antiferromagnetism, whereas it is a superconductor with incoherent spin density wave at larger doping. The presence of a QCP near optimal doping is also argued for hole--doped cuprates\cite{Varma1989}, but the situation appears somewhat different from the one of electron--doped cuprates: for the former, resistivity bears a signature of the QCP well above $T_c$ (see e.g. Ref.~\cite{Konstantinovic2001} and refs therein), and there is no evidence from the well documented ARPES data of a Fermi surface reconstruction at this point. 

In the case of LSCO, there is no study over a complete doping range that would allow to conclude on the existence of a similar QCP. However, if a similar QCP exists, doping by the field--effect is not expected to modify the Fermi surface in the same way as chemical doping. To fix the ideas, let us suppose a \textit{positive} gate voltage, promoting electron doping and hole depletion (Eq.~\ref{eq2}), and assume a Fermi surface as observed or computed for NCCO or PCCO (ref.). The increase and decrease for, respectively, the electron and hole density modifies the Fermi surface topology in the same way as chemical \textit{underdoping}. Thus, the increase in $T_c$ could not be in this case directly linked to the occurrence of a large reconstructed Fermi surface.

Then, superconductivity may be weakened on the underdoped side of the QCP, as a result of the quantum transition to a magnetic state: $T_c$ would decrease as the robustness of the antiferromagnetic states increases. Positive gate doping by the field--effect implies that the (hole) upper Hubbard band shifts up away from the Fermi level. However, the (electron) lower Hubbard band also shifts up, and the Hubbard gap may eventually decrease, weakening the antiferromagnetic state. Although this should be less efficient than chemical doping where both bands shift towards one another, this is a possible mechanism for the increase in $T_c$.

Finally, the increase of $T_c$ may be due to the increase in the superfluid density once the electrons are condensed (we assume that, as given by Eq.~\ref{eq2} and the above considerations on the respective magnitudes of electron and hole parameters, $\delta Q \approx \delta Q_e$). In Ref.~\onlinecite{Fruchter2010}, similar SLCO samples showed large value of the penetration depth; it was speculated that the superconducting transition is dominated by phase fluctuations (either of thermal or quantum origin). This was also suggested in Ref.~\onlinecite{Das2007}, observing that the experimental superfluid density for underdoped Nd$_{2-x}$Ce$_x$CuO$_4$ is well below the expectation from the computation of the area of the Fermi surface pockets, as a possible interaction of the antiferromagnetic fluctuations with superconductivity.
Again, the strong decrease of the superfluid density in the underdoped regime may be due to the presence of the magnetic phase, and thus be driven by the proximity of the QCP.

It is not, at present, possible to conclude on which of these mechanisms accounts for the observed $T_c$ variation. One limitation is the absence of the Fermi surface characterization, as well as the lack of a complete phase diagram ranging from underdoping to overdoping for the present material. Another limitation is the absence of a quantitative determination of the field--effect, which could be gained from ultra-thin films studies. It would thus be interesting to conduct similar experiments on other electron-doped materials: for overdoped states (which cannot be obtained at present for our material), we expect that the electric--field effect on $T_c$ should be ruled by the contributions of electron and holes, on a mean field basis, due to the absence of the phase fluctuation mechanism. If, as we believe, the contribution of the hole band to superconductivity is marginal, we then expect the electric--field effect to have a similar effect as chemical doping.

\newpage


\end{document}